\documentclass[twocolumn]{aastex62}

\newcommand{\GALEX}{{\it GALEX}}
\newcommand{\Gaia}{{\it Gaia}}

\newcommand{\HST}{{\it HST}}

\newcommand{\Spitzer}{{\it Spitzer}}
\newcommand{\Swift}{{\it Swift}}
\newcommand{\WISE}{{\it WISE}}

\newcommand{\Teff}{T_{\rm eff}}

\newcommand{\kms}{{\>\rm km\>s^{-1}}}

\newcommand{\bd}{BD+14$^\circ$3061}
\newcommand{\oCen}{$\omega$~Cen}
\newcommand{\uBVI}{{\it uBVI}}

\submitjournal{AJ}

\shorttitle{Post-AGB Stars in M19}
\shortauthors{Bond et al.}

\begin{document}          

\title{Two Luminous Post-AGB Stars in the Galactic Globular Cluster M19}

\correspondingauthor{Howard E. Bond}
\email{heb11@psu.edu}

\author[0000-0003-1377-7145]{Howard E. Bond}  
\affil{Department of Astronomy \& Astrophysics, Pennsylvania State
University, University Park, PA 16802, USA}
\affil{Space Telescope Science Institute, 3700 San Martin Drive,
Baltimore, MD 21218, USA}
\affil{Visiting Astronomer, Cerro Tololo Inter-American Observatory, National Optical Astronomy Observatory, operated by the Association of Universities for Research in Astronomy under a cooperative agreement with the National Science Foundation.}

\author[0000-0002-8994-6489]{Brian D. Davis}
\affil{Department of Astronomy \& Astrophysics, Pennsylvania
State University, University Park, PA 16802, USA}

\author[0000-0003-1817-3009]{Michael H. Siegel}
\affil{Department of Astronomy \& Astrophysics,
Pennsylvania
State University, University Park, PA 16802, USA}

\author[0000-0002-1328-0211]{Robin Ciardullo}
\affil{Department of Astronomy \& Astrophysics,
Pennsylvania
State University, University Park, PA 16802, USA}
\affil{Institute for Gravitation and the Cosmos, Pennsylvania
State University, University Park, PA 16802, USA}

\begin{abstract}

We report the discovery of a luminous ``yellow'' post-asymptotic-giant-branch (PAGB) star in the globular cluster (GC) M19 (NGC~6273), identified during our \uBVI\/ survey of Galactic GCs. The \uBVI\/ photometric system is optimized to detect stars with large Balmer discontinuities, indicating very low surface gravities and high luminosities. The spectral-energy distribution (SED) of the star is consistent with an effective temperature of about 6250~K and a surface gravity of $\log g=0.5$. We use \Gaia\/ data to show that the star's proper motion and radial velocity are consistent with cluster membership. One aim of our program is to test yellow PAGB stars as candidate Population~II standard candles for determining extragalactic distances. We derive a visual absolute magnitude of $M_V=-3.39\pm0.09$ for the M19 star. This is in close agreement with the $M_V$ values found for yellow PAGB stars in the GCs \oCen, NGC~5986, and M79, indicating a very narrow luminosity function. These objects are four magnitudes brighter than RR~Lyrae variables, and they can largely avoid the issues of interstellar extinction that are a problem for Population~I distance indicators. We also identified a second luminous PAGB object in M19, this one a hotter ``UV-bright'' star. Its SED is consistent with an effective temperature of about 11,750~K and $\log g=2.0$. The two objects have nearly identical bolometric luminosities, $\log L/L_\odot=3.24$ and 3.22, respectively.

\end{abstract}


\section{Yellow Post-AGB Stars in Globular Clusters}

The visually brightest objects in globular clusters (GCs) and old stellar
populations are stars in their final evolution from the tip of the
asymptotic giant branch (AGB) toward higher temperatures in the color-magnitude
diagram (CMD)\null. As these post-AGB (PAGB) stars pass at nearly constant
bolometric luminosity through spectral types early G, F, and late A, they reach
their brightest visual absolute magnitudes, because of the dependence of the
bolometric correction upon effective temperature.

It has been suggested \citep{Bond1997a, Bond1997b} that these luminous ``yellow'' PAGB stars
may be useful Population~II standard candles. In populations containing only old stars there should be fairly
sharp upper and lower limits to their brightnesses. In addition, yellow PAGB stars have
conspicuously large Balmer discontinuities in their spectral-energy
distributions (SEDs), making them easy to recognize using a single set of suitable photometric
observations. They should be detectable in early-type galaxies that do not
contain Cepheids, and in the halos of spirals, where there are fewer issues of
interstellar extinction than there are for Population~I distance indicators.

Because of the rapid evolutionary timescales for PAGB stars, however, these objects are very rare. The previously known luminous yellow PAGB stars in the Galactic GC system number only five: (1)~HD~116745 (``Fehrenbach's star,'' ROA~24; \citealt{Gonzalez1992} and references therein) and the RV~Tauri variable V1 in \oCen tauri \citep{Jones1968}; (2)~two luminous yellow PAGB stars in NGC~5986 \citep{Bond1977, Alves2001}; and (3)~a yellow PAGB star in M79, discovered by \citet[][hereafter B16]{Bond2016}. There are very few, if any, yellow stars in the CMDs of Galactic GCs that lie above the horizontal branch (HB) and within $\sim$1~magnitude of the brightness of these objects (see \citealt{Davis2021}, hereafter D21). Analogs of the GC objects are also known in the field; an example is the Galactic halo star \bd, along with a few other similar field objects (\citealt{Bond2020}, hereafter B20, and references therein).

In this paper, we report our discovery of a luminous yellow PAGB star in the
Galactic GC M19 (NGC~6273). As described in detail in D21, this object was the only new luminous yellow PAGB star found in our photometric survey of nearly the entire known sample of Galactic
GCs. Thus, it is likely to be the final member of this class in the Galactic GCs.  

As PAGB stars continue to evolve, they reach high effective temperatures, and
arrive at the top of the white-dwarf cooling sequence in the CMD\null. These
hot, luminous objects are conspicuous at short wavelengths, especially in the
ground- and space-based ultraviolet (UV). There is a substantial literature on
these prominent, but rare, ``UV-bright'' objects in GCs, beginning with \citet[][hereafter ZNG]{Zinn1972}, and summarized recently by \citet{Moehler2019}, who list about three dozen known or candidate members of the class. Our observations of M19 also revealed a previously unrecognized luminous UV-bright PAGB star, and we briefly discuss this object as well.


\section{Observations \label{sec:observations}}

In the mid-1990s, H.E.B. and collaborators began to develop a ground-based
photometric system optimized for efficient discovery of luminous, low-gravity A,
F, and G-type stars having large Balmer jumps. This
``{\it uBVI\/}'' system combines the \citet{Thuan1976} $u$ filter, whose
bandpass lies almost entirely shortward of the Balmer
discontinuity,\footnote{The Thuan-Gunn $u$ bandpass is distinct from, and
shortward of, that of the $u'$ filter of the Sloan Digital Sky Survey, which
partially overlaps the Balmer jump.} with the
broad-band {\it BVI\/} filters of the standard Johnson-Kron-Cousins system. The
astrophysical motivations and design principles of the {\it uBVI\/} system were
presented by \citet[][hereafter Paper~I]{Bond2005}. \citet[][hereafter Paper~II]{Siegel2005} established a network of equatorial \uBVI\/ standard stars, based on
extensive CCD observations with 0.9-, 1.5-, and 4-m telescopes at Kitt Peak
National Observatory (KPNO) and Cerro Tololo Inter-American Observatory (CTIO).

Over the years 1994 to 2001, H.E.B. made \uBVI\/ observations at KPNO and CTIO
of nearly the entire known sample of Galactic GCs. The primary aim of this
survey was to search systematically for yellow PAGB stars, with a goal of testing
their potential as standard candles and to establish a photometric zero-point.
A forthcoming paper (D21) will present a complete census of stars in the Galactic GC system lying above the HB, bluer than the red-giant branch and AGB, and redder than
$(B-V)_0=-0.1$ in the cluster CMDs. Further papers will describe the
zero-point calibration for yellow PAGB stars, and the results of searches for
them in Local Group galaxies. This paper focuses on the luminous yellow PAGB
star that the survey revealed in M19, as well as a luminous hot UV-bright member of the cluster.

\uBVI\/ survey observations of M19 were made with the CTIO 0.9-m telescope
on 1998 April~21 ($2\times2$ grid centered on the cluster; exposure times at each pointing were $2\times800$, 75, 40, and 45~s, respectively) and April~22 (cluster center; exposure times
400, 75, 40, and 45~s). The field of view of the 0.9-m CCD camera was
$13'\times13'$. Exposure times in the \uBVI\/ filters on the second night were
chosen so as to reach a signal-to-noise ratio of about 200--300 for stars about
2~mag brighter than the HB, allowing for interstellar extinction. The first
night was not photometric, and the exposure times in $u$ were lengthened.



M19 is a populous GC, one of the 10 or so most massive clusters in the Galactic GC system. \citet{Kruijssen2020}, \citet{Pfeffer2020}, and others have argued that it is the remnant nuclear star cluster of a dwarf galaxy (``Kraken''), which was disrupted and accreted by the Milky Way. M19 lies in the Galactic bulge in Ophiuchus ($l=356\fdg9, b=+9\fdg4$), and is overlain by a considerable number of field stars. Early work (e.g., \citealt{Harris1976}) showed that M19 is affected by substantial differential reddening, and that it has a very blue HB, consistent with low metallicity. 
The \citet[][hereafter H10]{Harris2010} catalog of GC parameters\footnote{Online version of 2010 December, at \url{http://physwww.mcmaster.ca/~harris/mwgc.dat}} gives a metal content of $\rm[Fe/H]=-1.74$. A recent spectroscopic study \citep{Johnson2017} of over 300 red giants and AGB stars in the cluster showed that there is a range of iron contents among the members, from about $\rm[Fe/H]=-2$ to $-1$, but concentrated around $-1.75$ and $-1.5$. These findings are indicative of considerable self-enrichment in this massive GC.

\section{Data Analysis and Selection of PAGB Candidates \label{sec:data}}

The CCD frames were reduced as described in detail by B16 and D21, using
standard tasks in IRAF\footnote{IRAF was distributed by the National Optical
Astronomy Observatory, operated by the Association of Universities for Research
in Astronomy (AURA) under a cooperative agreement with the National Science
Foundation.} for bias subtraction and flat-fielding. Instrumental stellar
magnitudes were measured on the frames with the {\tt ALLSTAR} and {\tt DAOGROW} tasks in
{\tt DAOPHOT} \citep{Stetson1987}. These data were corrected for atmospheric extinction,
and then the 1998 April~22 frames were calibrated to the \uBVI\/ system using
measurements of standard fields obtained during this observing run. We used the
$u$ magnitudes of standard stars from Paper~II, and $BVI$ magnitudes for the standards of
\citet{Landolt1992}. Since the 1998 April~21 night was not photometric, its
zero-points were determined by scaling to the M19 frames obtained on the 
photometric night. Finally, all of the measurements were combined into a single
catalog of mean calibrated magnitudes. For the bright PAGB candidates 
considered here, the photometry provides essentially complete stellar samples
into the cluster center (apart from rare cases of badly blended bright stars;
see D21 for further discussion of the sample completeness for our GC survey).

\begin{figure*}
\begin{center}
\includegraphics[width=2.5in]{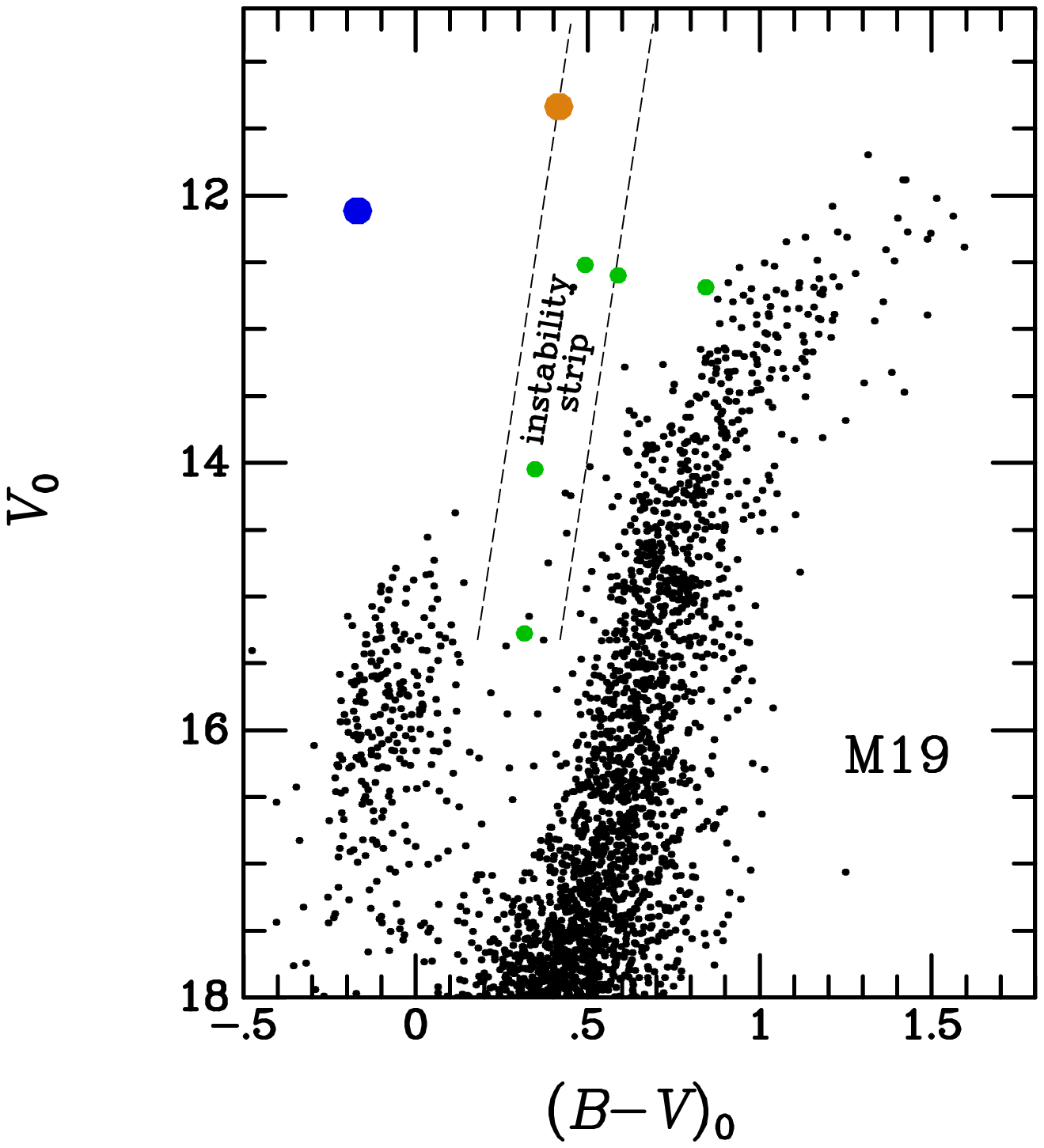}
\hfil
\includegraphics[width=2.5in]{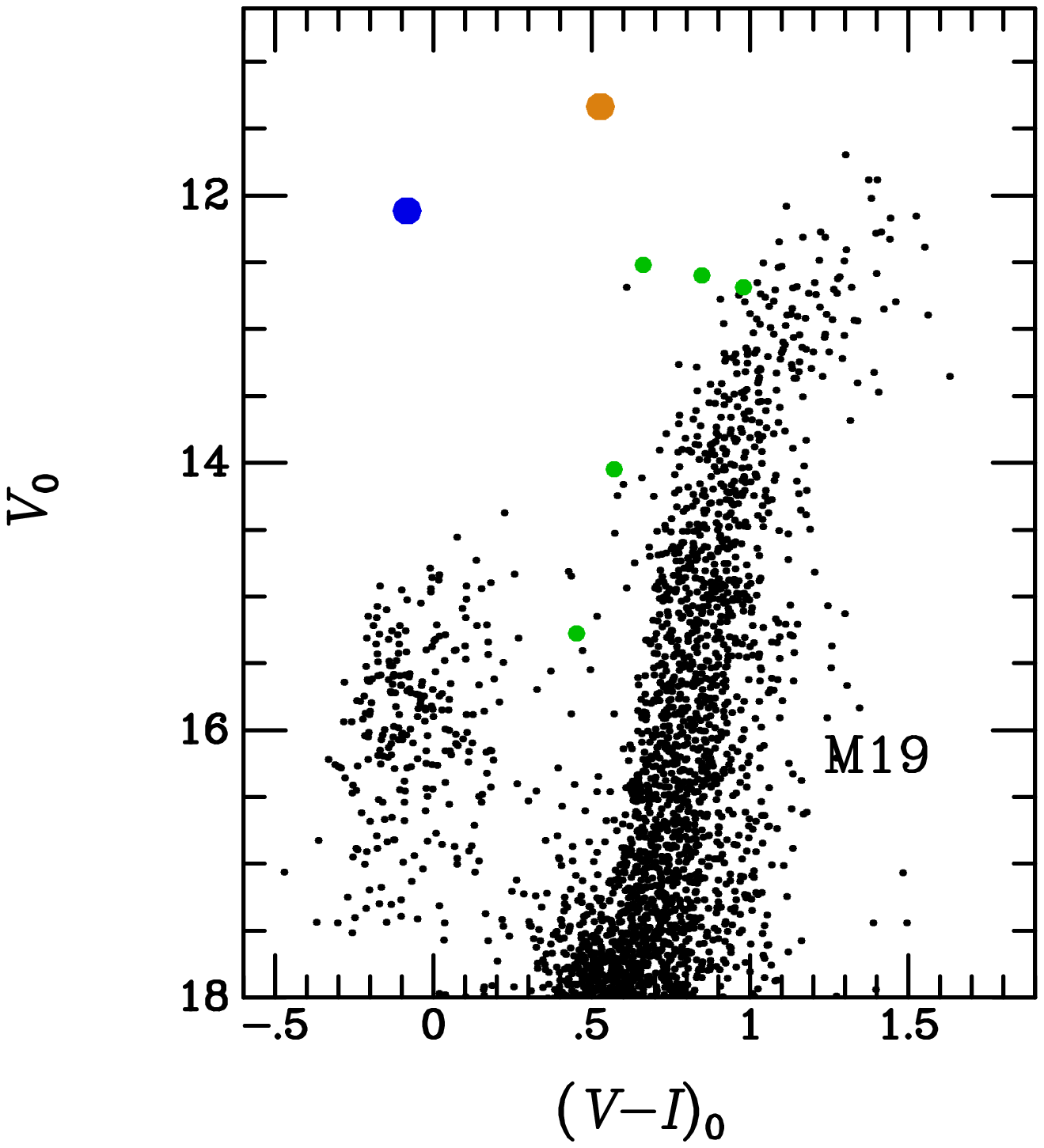}
\vskip0.2in
\includegraphics[width=3.5in]{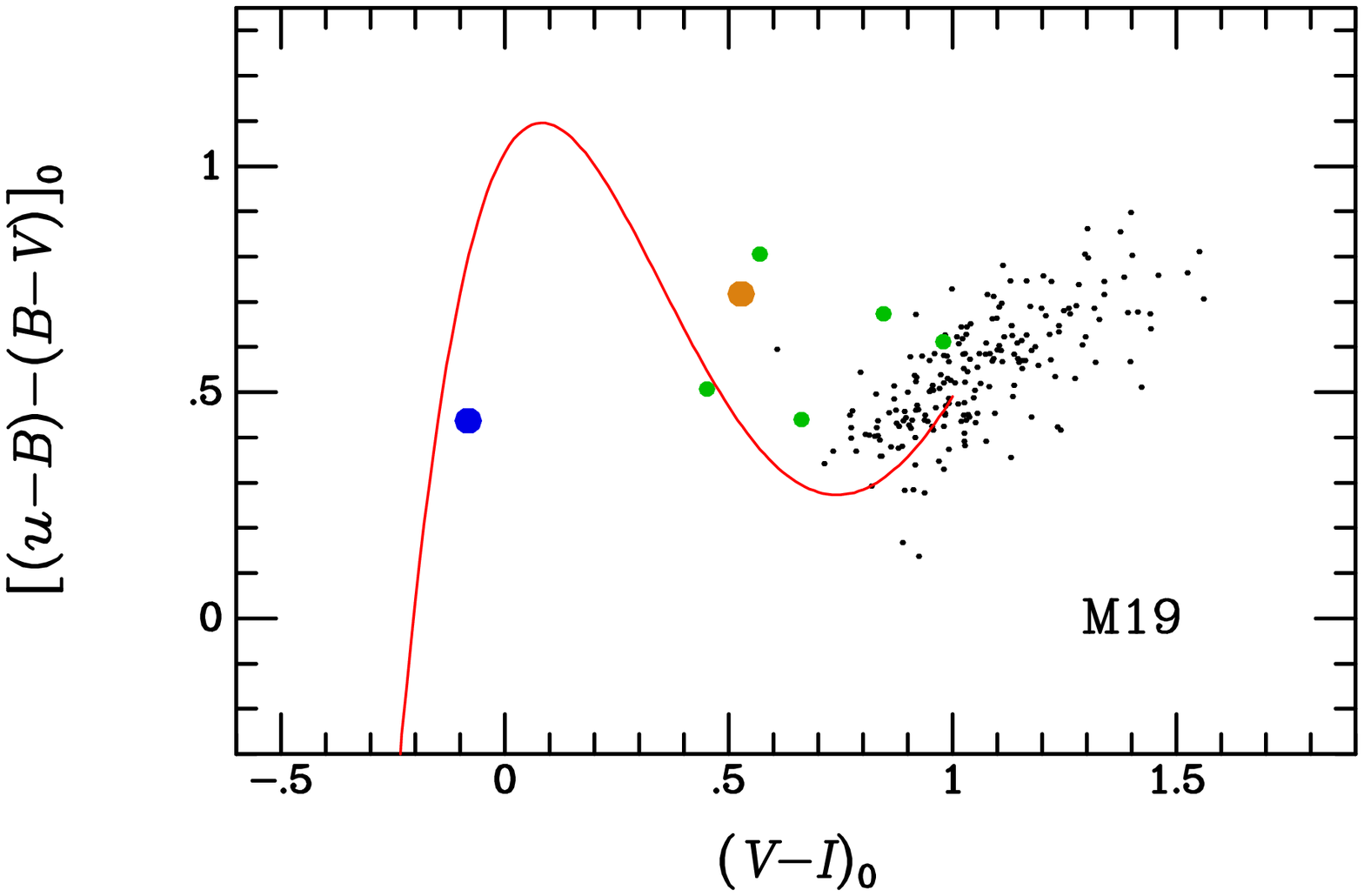}
\figcaption{\footnotesize
Top panels: Color-magnitude diagrams for M19 in $V,B-V$ and $V,V-I$\null. Our
photometry has been corrected for foreground reddening of $E(B-V)=0.38$, and
\Gaia\/ DR2 proper motions and parallaxes have been used to reject 
field stars (see text).  The filled orange and blue
circles show the locations of the two luminous PAGB stars belonging to the
cluster. Dashed lines indicate the location of the pulsational instability strip. Filled green circles mark one RR~Lyrae variable (the faintest object) and four Type~II Cepheids.
Bottom panel: Gravity-sensitive $(u-B)-(B-V)$ color index, corrected for
extinction and plotted against the temperature-sensitive $V-I$ color, for stars
brighter than $V_0=14$ (black points). The red curve is a polynomial fit to the location of the HB in a sample of GCs with redder HBs than M19's, taken from D21. The yellow PAGB star (filled orange circle), and the Cepheids, stand out from the HB locus because of their lower gravities and larger Balmer jumps. The blue PAGB star (filled blue circle) is hotter, and lies in the regime where the size of the Balmer jump depends on temperature rather than surface gravity. 
\label{fig:CMDs}}
\end{center}
\end{figure*}

We then searched for candidate yellow PAGB stars by choosing objects that are
simultaneously bright, have large Balmer jumps, and have spatial locations and
proper motions consistent with cluster membership. Figure~\ref{fig:CMDs} illustrates the
selection process. The top two frames show two CMDs for the cluster, after
correcting for a nominal cluster-averaged interstellar extinction of $E(B-V)=0.38$ (from H10),  and assuming $R_V=3.1$. (For the extinction corrections in $V-I$, we used the
formula of \citealt{Dean1978}.) The top left panel plots $V_0$ vs.\ $(B-V)_0$,
and the top right plots $V_0$ vs.\ $(V-I)_0$. 

As noted in \S\ref{sec:observations}, M19 lies in the Galactic bulge, and there
are substantial numbers of field stars in our CCD frames. To remove most of them
from our \uBVI\/ catalog, we applied three constraints, based on spatial location and on astrometry of the stars given in the \Gaia\/ Data Release~2 
(DR2; \citealt{Gaia2016, Gaia2018})\footnote{\url{http://vizier.cfa.harvard.edu/viz-bin/VizieR-3?-source=I/345/gaia2}}:
(1)~angular distance from the cluster center less than three times the half-light radius
(i.e., $3\times79\farcs2$, using the half-light radius given by H10); (2)~DR2
proper motions in right ascension and declination within $1.1\rm\,mas\,yr^{-1}$
of the cluster mean of $(\mu_\alpha,\mu_\delta)=(-3.22,
+1.61)\rm\,mas\,yr^{-1}$ \citep{GaiaGC2018}; and (3)~DR2 parallax less than 0.7~mas (to exclude objects almost certainly in the foreground). The top panels
in Figure~\ref{fig:CMDs} show that these criteria result in a nearly pure sample of cluster
members, especially brighter than $V_0\simeq14$; but there do appear to be a few
remaining field stars below that level.





Also shown in the top left panel is the approximate location of the RR~Lyrae,
Cepheid, and RV~Tauri instability strip (e.g., \citealt{Harris1983}; D21). M19 contains seven
objects listed in the \citet{Clement2001} catalog\footnote{Updated version available online at \url{http://www.astro.utoronto.ca/~cclement/cat/listngc.html}} of variable stars in GCs. However, two of them (V6 and V7) appear to be non-members, based on their proper motions in \Gaia\/ DR2. The positions of the remaining five variables in the CMDs in Figure~\ref{fig:CMDs}, based on our photometry, are plotted as filled green circles. They consist of four Type~II Cepheids\footnote{\citet{Clement1978} point out that the only Galactic GCs having more Type~II Cepheids than M19 are \oCen\ and M14.} and one RR~Lyrae variable. Our data are comprised of only a few measurements on two successive nights, made at random pulsation phases, so they should not be regarded as representing the time-averaged locations of the
variables in the CMDs. Nevertheless, the RR~Lyr variable, and three of the
Cepheids, do lie inside, or on the border of, the schematic instability strip.

The red-giant branch (RGB) and HB of the cluster's CMD in Figure~\ref{fig:CMDs} have
appreciable width (compare with, for example, the very narrow RGB and HB of the
nearly unreddened GC M79 in our team's data, shown in Figure~1 of B16). This
spread is partially due to substantial differential interstellar reddening across the face of M19---as we noted above, and has been discussed in several more recent studies of the cluster,
including \citet{Piotto1999}, \citet{Alonso2012}, \citet{Johnson2017},
and references therein to earlier literature.

As our index of the strength of the Balmer discontinuity, we use the color
difference $(u-B)-(B-V)$, which is a broad-band analog of the $c_1$ Balmer-jump index of
the Str\"omgren $uvby$ system.\footnote{Use of a color difference has the
advantage of only a weak dependence on interstellar extinction; a formula for
the extinction correction as a function of $E(B-V)$ is given in Paper~I.} In the
bottom panel of Figure~\ref{fig:CMDs}, we plot this color difference vs.\ the $(V-I)_0$
temperature index, for the M19 members in the top panels that are brighter than
$V_0=14.0$. At these bright magnitudes, nearly all of the M19 stars are red
giants or AGB stars. The selection criterion for yellow PAGB stars is
that they have a Balmer-jump index significantly larger (i.e., redder) than do
HB stars at the same color, and that they are at least 3~magnitudes brighter than
the HB at the same color. As the top panels show, the HB in M19 is extremely blue, and almost
entirely lacks stars on the cooler ``horizontal'' locus of the HB\null.
Therefore, as an indicator of the HB's location, we use a polynomial fit to the
mean location of the HB in a sample of GCs with redder HBs, taken from D21. This
relation is plotted as a red line in the bottom panel of Figure~\ref{fig:CMDs}.
The peak in this sequence on the left-hand side of the
diagram, at $(V-I)_0\simeq0.1$, is due to the maximum size of the Balmer jump at this color on the HB.  

The single RR~Lyrae star in M19 (included in the bottom panel of Figure~\ref{fig:CMDs} in
spite of being fainter than $V_0=14.0$) does lie almost exactly on the mean HB
locus. The four Cepheids fall above the HB relation, consistent with their
higher luminosities, lower surface gravities, and consequent higher
$(u-B)-(B-V)$ color differences. There also appears to be an above-the-HB (AHB)
member of M19 at $V_0\simeq12.7$ with a high value of the color difference;
curiously, although this star apparently lies within the instability strip, it
is not a known variable. This object, and other AHB stars in Galactic GCs, will
be discussed in D21.

The most conspicuous star in the three panels of Figure~\ref{fig:CMDs} is the very bright
object plotted as an orange filled circle. It has a Balmer jump larger than
those of HB stars, as shown in the bottom panel, making it a strong candidate
for a yellow PAGB star.

Also conspicuous in the Figure~\ref{fig:CMDs} CMD is a bright, very blue star. For stars this
hot, the $(u-B)-(B-V)$ color difference is no longer sensitive to surface
gravity (see Paper~I, Figures~4 and~5), becoming instead a temperature index. This
star is a candidate luminous blue PAGB star.

Both objects had been marked as candidate UV-bright stars in the
classical ZNG study; our yellow PAGB star is M19 ZNG~4, and the blue one
is M19 ZNG~2. To our knowledge, the present study is the first to present evidence that both of them are luminous members of the cluster, rather than foreground objects.

\begin{figure*}[hbt]
\begin{center}
\includegraphics[height=2.7in]{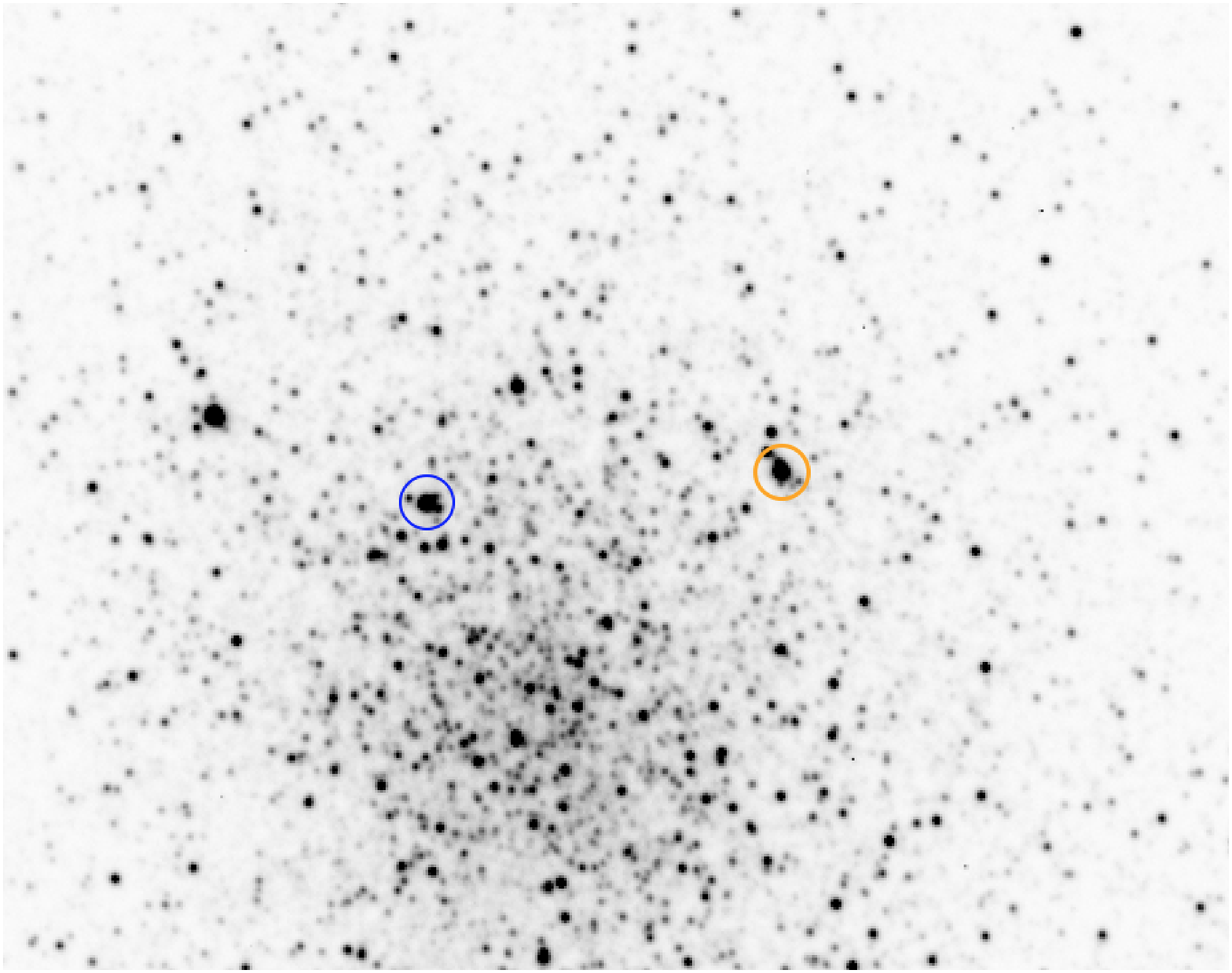}
\hfil
\includegraphics[height=2.7in]{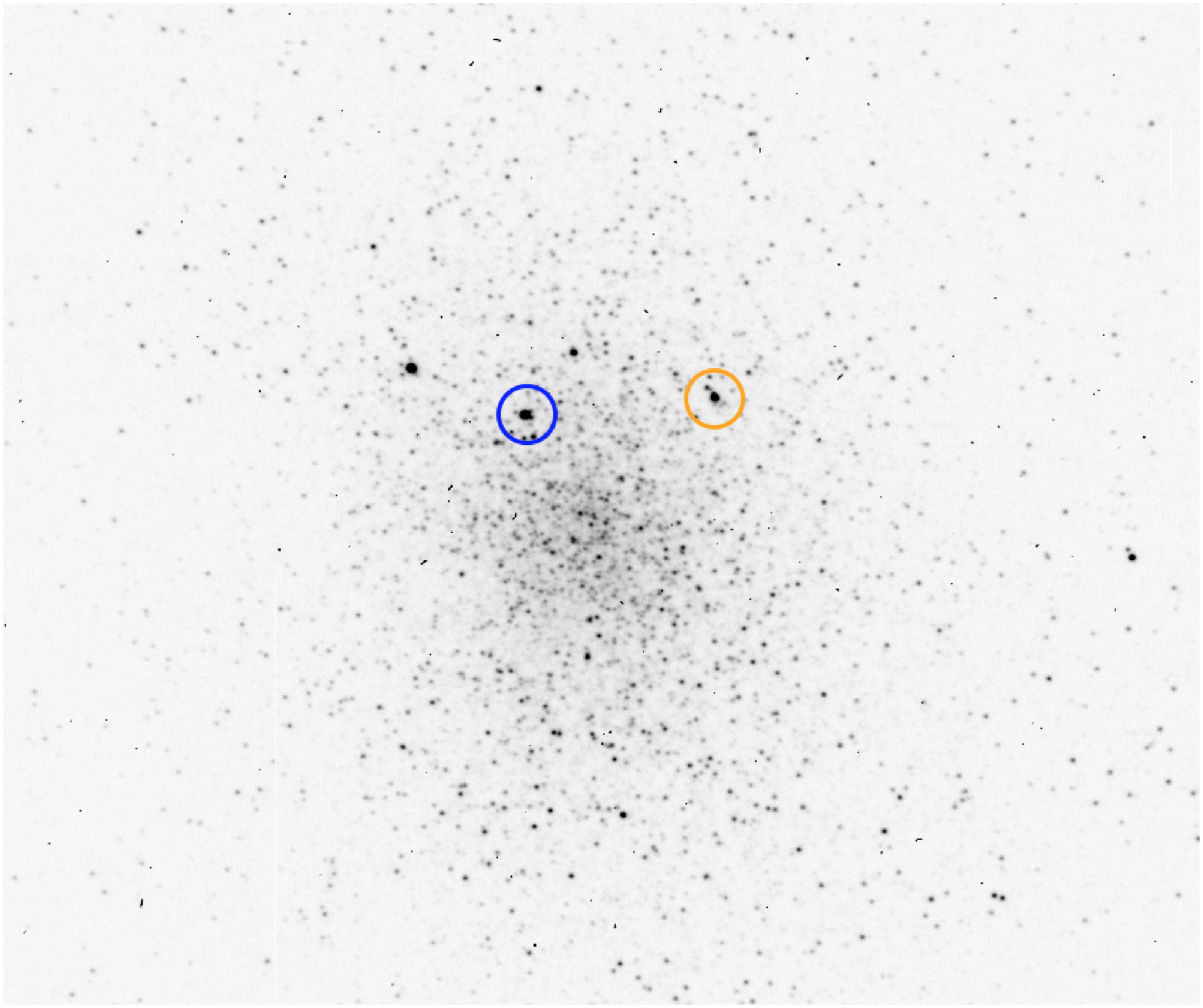}
\figcaption{\footnotesize
Left panel: finding chart for the yellow PAGB star (orange circle) and the blue
PAGB star (blue circle) in the globular cluster M19. Made from a $B$-band CCD
frame obtained with the CTIO 0.9-m telescope. North is at the top, east on the
left; frame is $2\farcm5$ high. 
Right panel: $u$-band frame from same telescope, height $4\farcm8$,
illustrating how conspicuous both stars are in the
ground-based ultraviolet. ({ The bright star to the NE of the blue
PAGB star is a foreground object, ZNG~1. The bright star to the NW is the Type~II Cepheid V1 = ZNG~3.})
\label{fig:charts}
}
\end{center}
\end{figure*}

The left panel in Figure~2 presents a finding chart for the yellow and blue PAGB
stars in M19, made from one of our $B$-band frames. To illustrate how conspicuous
these two stars are in the ground-based UV, the right panel in
Figure~\ref{fig:charts} shows a $u$-band frame. The two PAGB stars dominate this image. ({The bright star to the NW of the blue PAGB star is the brightest Type~II Cepheid in M19, V1, which was marked as the UV-bright object ZNG~3 in the ZNG study. The 
bright star to the NE of the blue PAGB star is ZNG~1, a foreground field
object, as indicated by its \Gaia\/ parallax and proper motion. Two more UV-bright candidates, further from the cluster center, designated ZNG~5 and 6, are also non-members, according to \Gaia\/ astrometry.})  

Table~\ref{table:basicdata} presents basic data for our two PAGB candidates. The \Gaia\/ Early Data Release~3 (EDR3; \citealt{Gaia2020}) became available as we were completing this paper, and is the source for the astrometry given in rows 1 through 5 of the table. The radial velocity (RV) for the yellow PAGB star in row 6 is from \Gaia\/ DR2. Rows 7 through 10 give our \uBVI\/ photometry. The
interstellar reddenings in row~11 are taken from the high-resolution M19
extinction maps of \citet{Alonso2012} and \citet{Johnson2017}. The
former map is based on stellar photometry with the Magellan Telescope, and the
latter on stellar photometry using {\it Hubble Space Telescope\/} (\HST)
images.\footnote{The Alonso-Garcia et al.\ reddening map is presented
pictorially in their Figure~13, and in tabular form at
\url{https://vizier.u-strasbg.fr/viz-bin/VizieR-4}. C.~Johnson kindly sent us a
numerical table giving the extinction map depicted in Figure~2 of \citet{Johnson2017}. Unfortunately the existing \HST\/ frames cannot be used for optical photometry of the
bright PAGB stars because their images are saturated.} The reddening values from
these two maps agree well [the differences in $E(B-V)$ at the locations of the two stars are 0.006 and 0.013 mag,
respectively], and we give their means in Table~\ref{table:basicdata}. However, it should be noted
that there are appreciable variations in $E(B-V)$ from cell to cell in both
extinction maps, and it is difficult to estimate the uncertainty in the
reddening for an individual star. We adopt $\pm$0.02 mag as a reasonable guess.

\begin{deluxetable*}{lccc}[ht!]
\tablewidth{0 pt}
\tablecaption{Basic Data for the Luminous Post-AGB Stars in M19
\label{table:basicdata}
}
\tablehead{
\colhead{Parameter} & 
\colhead{``Yellow'' PAGB} &
\colhead{``Blue'' PAGB} &
\colhead{Source\tablenotemark{a}} \\
\colhead{} & 
\colhead{(M19 ZNG~4)} &
\colhead{(M19 ZNG~2)} &
\colhead{} 
}
\startdata
R.A. [J2000]     & 17:02:35.186   & 17:02:39.155 & (1) \\
Dec. [J2000]     & $-26$:15:24.14   & $-26$:15:29.36& (1) \\
Parallax [mas]        & $+0.119\pm0.016$ & $+0.158\pm0.024$ &(1) \\
R.A. proper motion [$\rm\,mas\,yr^{-1}$] & $-2.878\pm0.019$ & $-2.990\pm0.025$ &(1) \\
Dec. proper motion [$\rm\,mas\,yr^{-1}$] & $+1.146\pm0.012$ & $+1.454\pm0.018$ &(1) \\
Radial velocity [$\kms$] & $+142.3\pm1.4$ & $\dots$ & (2) \\
$V$   & $12.512\pm0.006$ & $13.291\pm0.006$ & (3) \\
$u-B$ & $1.471\pm0.012$  & $0.608\pm0.015$  & (3) \\
$B-V$ & $0.795\pm0.010$  & $0.211\pm0.010$  & (3) \\
$V-I$ & $1.021\pm0.011$  & $0.390\pm0.010$  & (3) \\
Reddening, $E(B-V)$ & $0.344\pm0.020$ & $0.342\pm0.020$ & (4) \\
Absolute magnitude, $M_V$ & $-3.39\pm0.09$ & $-2.61\pm0.09$ & (5) \\
Absolute luminosity, $\log L/L_\odot$  & 3.24           & 3.22           & (6)
\enddata
\tablenotetext{a}{Sources: (1)~\Gaia\/ EDR3; (2)~\Gaia\/ DR2; (3)~This paper (not corrected for extinction); note that the zero-point for the $u$ magnitude is such that $u=1.0$ for Vega; (4)~From the reddening maps of \citet{Alonso2012} and \citet{Johnson2017} (see the text, \S\ref{sec:data}); (5)~This paper, calculated from data in this table, $R_V=3.1$, and a distance modulus of $(m-M)_0=14.84\pm0.06$~mag (see the text, \S\ref{sec:absmag}); (6)~This paper, calculated from data in this table, and bolometric corrections from \citet{Castelli2004} (see the text, \S\ref{sec:seds}).}  
\end{deluxetable*}

\bigbreak

\section{Cluster Distance and Visual Absolute Magnitudes \label{sec:absmag}}

To convert the apparent magnitudes in row~7 of Table~\ref{table:basicdata} to absolute magnitudes, we need the distance to the cluster. For M19, distance determinations from photometric methods are unusually problematic, for at least two reasons. First, the relatively large and spatially variable reddening makes the correction for extinction difficult. Second, distance methods based on the HB suffer because, in the case of M19 (as shown in Figure~\ref{fig:CMDs}), the cluster's ``horizontal'' branch is nearly vertical in CMDs made at optical wavelengths.

With the recent availability of \Gaia\/ EDR3, it is also possible to obtain a direct geometric distance estimate. We considered the sample of 202 RGB cluster members brighter than $V_0=14$, which we had chosen for the bottom panel in Figure~\ref{fig:CMDs}. The mean EDR3 parallax for this sample, and its standard error, are $0.0934\pm0.0062$~mas. To this we applied the global zero-point offset of +0.017~mas from the analysis of \citet{Lindegren2020}.

In Table~\ref{table:distance} we list six distance determinations, made using a variety of essentially independent techniques, as summarized in the notes to the table. We see no compelling reason to adopt one of these determinations, so we will use a weighted mean of the four determinations for which uncertainties are given (and which, as it happens, span the range of distances found by these studies). This results in our adopted distance modulus of $(m-M)_0=14.84\pm0.06$ ($d=9.29\pm0.26$~kpc). 

Row 12 in Table~\ref{table:basicdata} gives the visual absolute magnitudes of both stars,
calculated from our photometry, the interstellar extinction given in row~11
(using $R_V=3.1$), and the distance modulus adopted in this section. The stated formal uncertainties are likely somewhat optimistic, given the possibilities of systematic errors in the extinction and distance.

\begin{deluxetable}{ccl}[htb]
\tablewidth{0 pt}
\tablecaption{M19 Distance Determinations
\label{table:distance}
}
\tablehead{
\colhead{Distance [kpc]} & 
\colhead{$(m-M)_0$} &
\colhead{Source} 
}
\startdata
$8.99\pm0.83$   & $14.77\pm0.20$ &  \citet{Piotto1999}\tablenotemark{a} \\
$9.95\pm0.37$ & $14.99\pm0.08$ &   \citet{Recio2005}\tablenotemark{b} \\
8.24            & 14.58 &  \citet{Valenti2007}\tablenotemark{c} \\
8.79            & 14.72 &  \citet{Harris2010}\tablenotemark{d} \\
$8.13\pm0.47$ & $14.55\pm0.13$ & \citet{Baumgardt2019}\tablenotemark{e} \\
$9.06\pm0.51$   & $14.78\pm0.12$ & This paper, \Gaia\/ EDR3\tablenotemark{f} \\
\enddata
\tablenotetext{a}{Distance determined from luminosity of red-giant bump in \HST-based $BV$ color-magnitude diagram.} 
\tablenotetext{b}{Distance calculated by us by subtracting $3.1\,E(B-V)$ from the value of $(m-M)_{\rm F555W}=16.04\pm0.08$, and adopting $E(B-V)=0.34$, as given by \citet{Recio2005}.}  
\tablenotetext{c}{Distance determined from near-IR photometry of stars on the red-giant branch. Uncertainty not given.}
\tablenotetext{d}{Distance calculated by us by subtracting $3.1\,E(B-V)$ from the value of $(m-M)_V=15.90$, and adopting $E(B-V)=0.38$, as given by \citet{Harris2010}. Uncertainty not given.}
\tablenotetext{e}{``Kinematic'' distance derived from cluster-member proper-motion and radial-velocity dispersions and $N$-body modeling.}  
\tablenotetext{f}{Calculated by us based on the mean \Gaia\/ EDR3 parallax of a sample of cluster red giants, as described in \S\ref{sec:absmag}.}
\end{deluxetable}

\bigbreak \bigskip

\section{Cluster-Membership Tests}

Both of the PAGB candidates lie well within the spatial boundaries of the cluster (see Figure~\ref{fig:charts}): the yellow and blue stars fall $50''$ and $36''$ from the cluster center, respectively, which are within the half-light radius of $79\farcs2$. Moreover, they have \uBVI\/ colors that are extremely unusual for field stars. Thus it is already highly probable that they are members of the
cluster.

Two further membership tests are possible: (1)~RV, and (2)~proper
motion. Parallax is a less useful criterion for individual stars  at the large adopted distance of M19, which
corresponds to a parallax of only 0.108~mas. Nevertheless, the \Gaia\/ EDR3 parallaxes for
both stars, given in row~3 of Table~\ref{table:basicdata}, corrected for the +0.017 zero-point offset \citep{Lindegren2020}, agree with the nominal value to within 1.8 and 2.8$\sigma$, respectively. 

\Gaia\/ EDR3 did not give RVs for either star, but the earlier DR2 listed a RV of $+142.3\pm1.4\,\kms$ for the yellow PAGB
star. This agrees extremely well with the mean cluster RV of
$+145.54\pm0.59\,\kms$ \citep{Baumgardt2019}; the velocity dispersion in the cluster is $11.0\kms$ \citep{Baumgardt2018}. \Gaia\/ DR2 and EDR3 did not list a RV
for the early-type blue PAGB candidate.

\begin{figure*}[hbt]
\begin{center}
\includegraphics[width=3.0in]{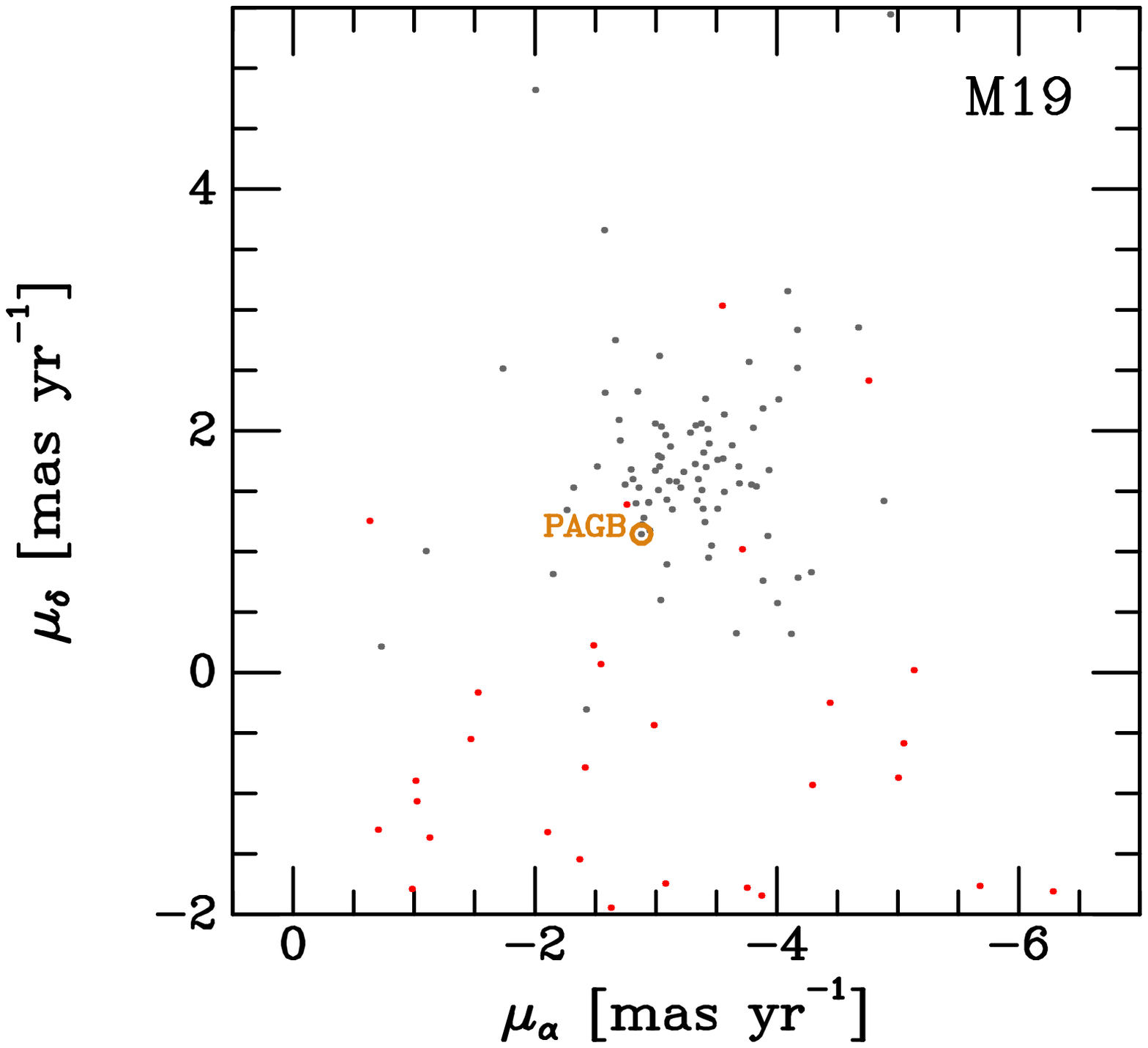}
\hfil
\includegraphics[width=3.0in]{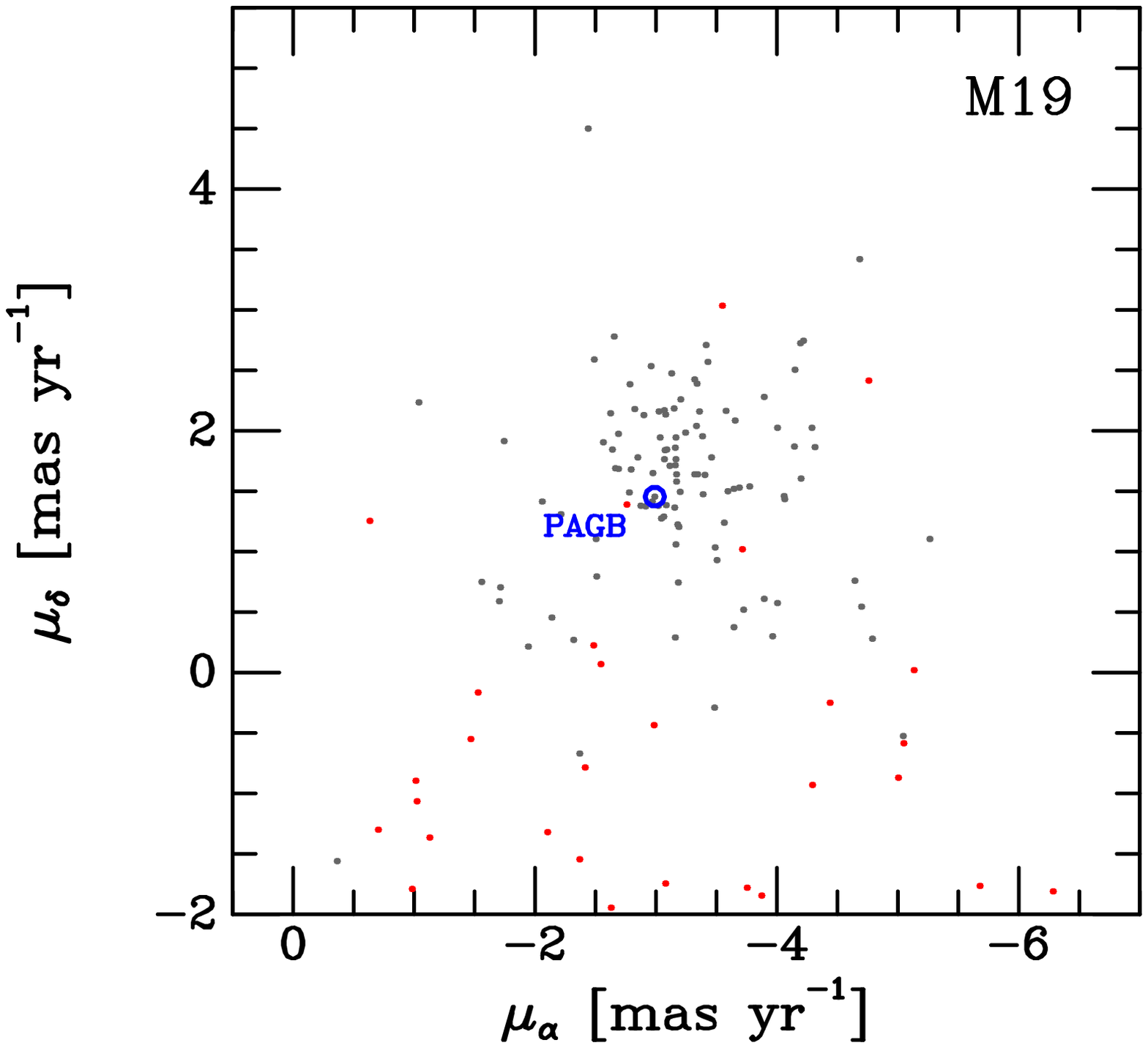}
\figcaption{\footnotesize
Left panel: Proper motions from \Gaia\/ EDR3 for stars within $30''$ of the yellow PAGB star in M19, with parallax less than 1~mas, brighter than $G=18$, and $BP-RP$ color redder than 1.2 (black points). The color-magnitude diagram of these stars (not shown) indicates that nearly all of them are cluster members on the red-giant branch. The proper motion of the yellow PAGB star is marked with an orange circle.  
Right panel: Proper motions from EDR3 for stars within $30''$ of the blue PAGB star in M19, selected with the same criteria (black points). The proper motion of the blue PAGB star is marked with a blue circle. In both panels the red points show the proper motions of stars in a nearby field, selected using the same criteria. The proper motions of both stars are well within the distributions of cluster members. 
\label{fig:pm}
}
\end{center}
\end{figure*}

To make proper-motion membership tests for the two stars, we selected nearly pure samples of M19 members from
\Gaia\/ EDR3. We chose stars lying within $30''$ of each object, brighter than magnitude
$G=18$, redder than $BP-RP=1.2$, and having a parallax less than 1~mas. The proper motions for these
samples are plotted as black points in the two panels in Figure~\ref{fig:pm}, with the proper motions of
the PAGB candidates themselves (rows~4 and 5 in Table~\ref{table:basicdata}) marked with orange and
blue open circles. The \Gaia\/ CMDs of these two samples indicate that a large
majority of the stars are cluster members lying on the RGB\null. In both cases, the motions of the
PAGB candidates are well within the distributions of the cluster members. To illustrate the proper-motion distribution in the surrounding field, we selected stars from EDR3 in a nearby field, with a radius of $200''$ and the same criteria. These are plotted as red points in the two figures; they show a wide range of proper motions, most of them in fact outside the range plotted.

In summary, the available evidence strongly confirms that both stars are
physical members of M19.

\section{Variability}

As shown in the top left panel in Figure~\ref{fig:CMDs}, the M19 yellow PAGB star lies close
to the instability strip in the CMD\null. Several field analogs of the yellow
PAGB stars in GCs are known to be low-amplitude semi-regular variables,
including HD~46703 and \bd. As discussed in B20 (and references therein), both
field stars have typical pulsation periods of $\sim$29--32~days, and
peak-to-peak amplitudes of $\sim$0.1--0.3~mag. The pulsation amplitudes are variable; in
the case of \bd, the variations can actually drop below detectability for
extended intervals.

The most recent extensive search for variable stars in M19 of which we are aware
is the photographic study by \citet{Clement1978}; this investigation
did not mark the yellow PAGB star as variable. Our own \uBVI\/ data are of
limited value, since we only have the two epochs in 1998 April separated by one
day, plus an earlier lower-quality CCD observation obtained with the CTIO 1.5-m
in 1995, which we did not include in our calibrated photometric reductions. We
see no convincing evidence for variability in this limited material at a level
of more than a few hundredths of a magnitude. The available data from current
all-sky monitoring programs with small telescopes are generally of limited use,
because of the crowding within the cluster. The uncertainty given for the $G$ magnitude in \Gaia\/ EDR3 is very small, consistent with little or no variability.

Thus there is no evidence that the yellow PAGB star is variable, but additional
monitoring observations with sufficient spatial resolution would be useful.
(There is also no evidence that the blue PAGB star is variable, based on the
same material.) 

\section{Spectral-Energy Distributions \label{sec:seds}}

We determined SEDs for both PAGB stars by combining our optical photometry from Table~\ref{table:basicdata} with public data from the following sources: 

(1)~The {\it Galaxy Evolution Explorer\/} (\GALEX) \citep{Morrissey2007} imaged M19 in its all-sky UV survey,\footnote{\url{https://galex.stsci.edu/GalexView/}} but only in its far-UV (FUV) bandpass. Although our blue PAGB star is prominent at FUV wavelengths, it is not contained in the \GALEX\/ source catalog (presumably because of crowding near the cluster center). 
We obtained the FUV image and, because the blue PAGB
is so much brighter than its neighboring objects at FUV wavelengths, we measured its
brightness using simple aperture photometry.  We then translated these raw magnitudes
into the standard flux system by scaling our measurements to similar aperture
photometry of nearby isolated field stars with magnitudes given in the \GALEX\/
source catalog.
The yellow PAGB star was too faint and blended to be measured in the FUV. 

(2)~The {\it Hubble Space Telescope\/} (\HST) has imaged M19 on a few occasions,\footnote{\url{https://archive.stsci.edu/}} but in nearly all of the frames the images of both PAGB stars are heavily saturated. We found an exposure of the blue PAGB star obtained with the Wide Field Planetary Camera~2 (WFPC2) in the near-UV (NUV) F255W filter (program GO-8718; PI G.~Piotto), in which only the central pixel was saturated; we performed aperture photometry to obtain a lower limit on the star's flux. For the yellow PAGB star, there are 10 WFPC2 frames in F255W (GO-10815; PI T.~Brown) with unsaturated images, on which we also performed aperture photometry. We used the {\tt PHOTFLAM} keyword in the image headers to convert the counts to absolute fluxes, and applied the standard 0.10~mag correction to an infinite aperture. The \HST\/ images are virtually unaffected by source crowding. 

(3)~The UV/Optical Telescope (UVOT) on the Neil Gehrels {\it Swift\/} Observatory obtained images\footnote{\url{https://archive.stsci.edu/swiftuvot/}} of M19 in 2009 and 2010 in its three NUV/FUV bandpasses, $uvw1$, $uvm2$, and $uvw2$. We combined these images, and then performed PSF-fitting photometry of both PAGB stars using {\tt DAOPHOT}\null. Corrections for time-dependent sensitivity loss, coincidence losses, and exposure time were applied, as detailed in \citet{Siegel2014}; the data were then calibrated to the absolute photometric system of \citet{Breeveld2011}. The $uvw1$ and $uvw2$ filters have substantial red leaks, which are problematic for cool stars; because of this, exacerbated by the considerable interstellar reddening, we did not include the observations of the yellow PAGB star in these two bandpasses in our analysis. 

(4)~The 2MASS near-infrared (NIR) sky survey \citep{Skrutskie2006} obtained images of M19 in $J$, $H$, and $K_s$. Photometry of our PAGB stars is contained in the 2MASS source catalog,\footnote{\url{https://irsa.ipac.caltech.edu/cgi-bin/Gator/nph-dd}} but is clearly affected by the stellar crowding near the cluster center. We therefore downloaded the 2MASS frames, and performed {\tt DAOPHOT} PSF photometry on the images, calibrating to the 2MASS photometric zero-point using measurements of nearby isolated stars contained in the source catalog. \HST\/ images in the $I$ (F814W) band show that the yellow PAGB star, at the 2MASS resolution, is blended with four nearby red giants. We attempted to de-blend the images, using the \HST\/ frame as a prior to define the precise locations of the PAGB stars in the 2MASS frames, but with limited success.

(5)~Images of M19 were obtained in 2013 with the warm {\it Spitzer Space Telescope\/} and its Infrared Array Camera (IRAC), in the 3.6 and $4.5\,\mu$m bandpasses; the observations were part of a program aimed at RR~Lyrae stars (PI W.~Freedman). We downloaded a selection of these images\footnote{\url{https://sha.ipac.caltech.edu/applications/Spitzer/SHA/}} and performed {\tt DAOPHOT} photometry to de-blend the stellar images, using PSF fitting radii of the order of the FWHM\null.
Aperture corrections were then applied, based on the total magnitudes of isolated
field stars in the frames.

(6)~We examined NIR and far-IR images of M19 from the {\it Wide-Field Infrared Survey Explorer\/} (\WISE\/) sky survey, which we downloaded from SkyView.\footnote{\url{https://skyview.gsfc.nasa.gov/current/cgi/query.pl}} However, the spatial resolution of these frames is far too poor to provide useful photometry for our targets lying in very crowded regions of the cluster.

We then corrected all of the measured fluxes for interstellar extinction, using the values of $E(B-V)$ in Table~\ref{table:basicdata}, and for most bandpasses applying the formulae of \citet{Cardelli1989}, calculated at the effective wavelength of each bandpass, and assuming $R_V=3.1$. For the UVOT photometry, because of the complex structure of the extinction curve around 2200~\AA, we determined the extinction corrections by integrating the Cardelli et al.\ formula convolved with the system throughput curves and a blackbody function having the approximate temperature of each star. For the two IRAC bandpasses, whose wavelengths are beyond the range of the Cardelli et al.\ formulae, we determined the extinction at the $K_s$ band, and scaled it to [3.6] and [4.5] according to the relations given by \citet{Indebetouw2005}. The resulting extinction-corrected fluxes must be regarded as only approximate, given the very large correction factors in the UV, the uncertainty whether $R_V=3.1$ is appropriate for this line of sight, and the source crowding in many of the bandpasses. 

\begin{figure*}[hbt]
\begin{center}
\includegraphics[width=3.3in]{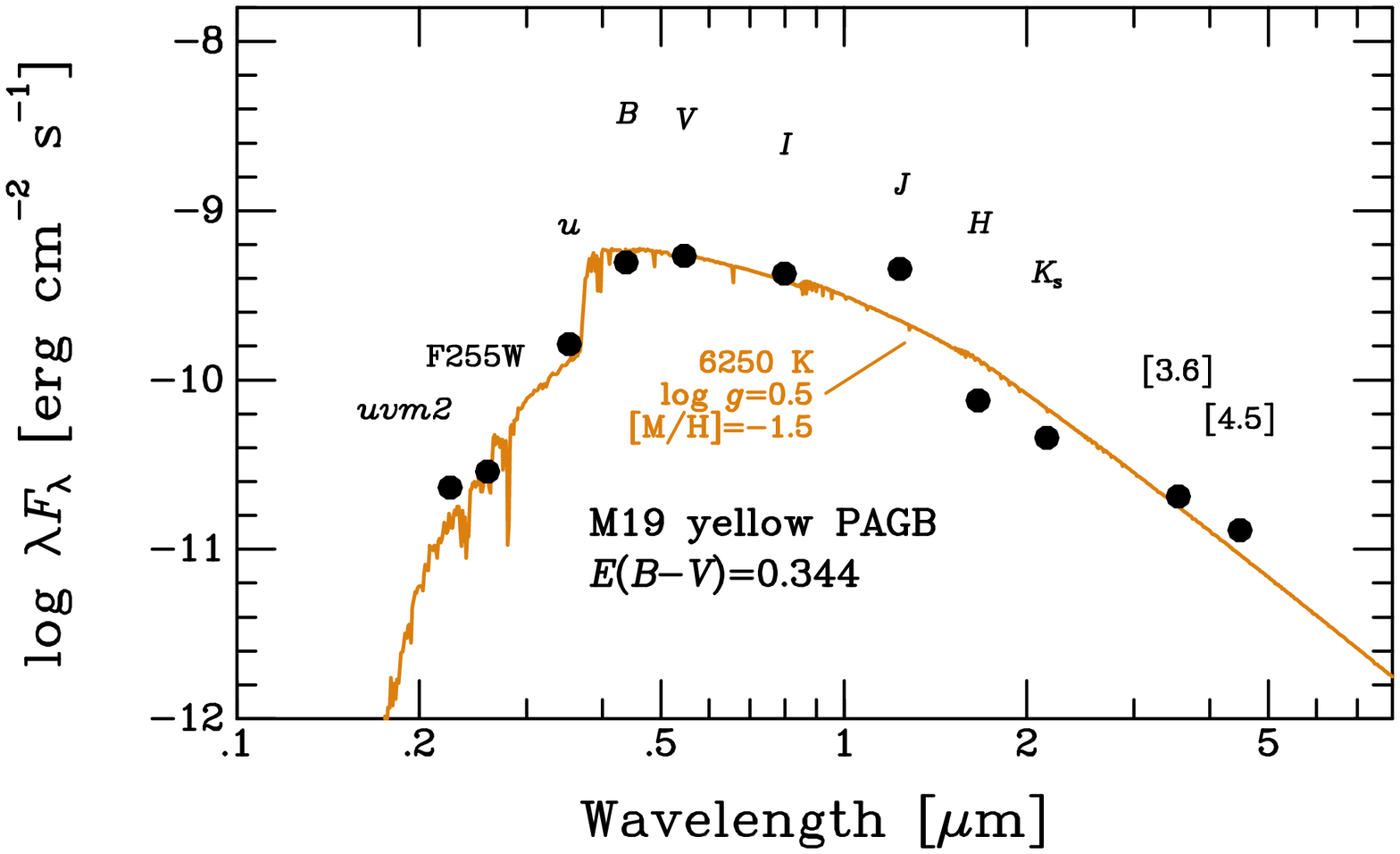}
\hfill
\includegraphics[width=3.3in]{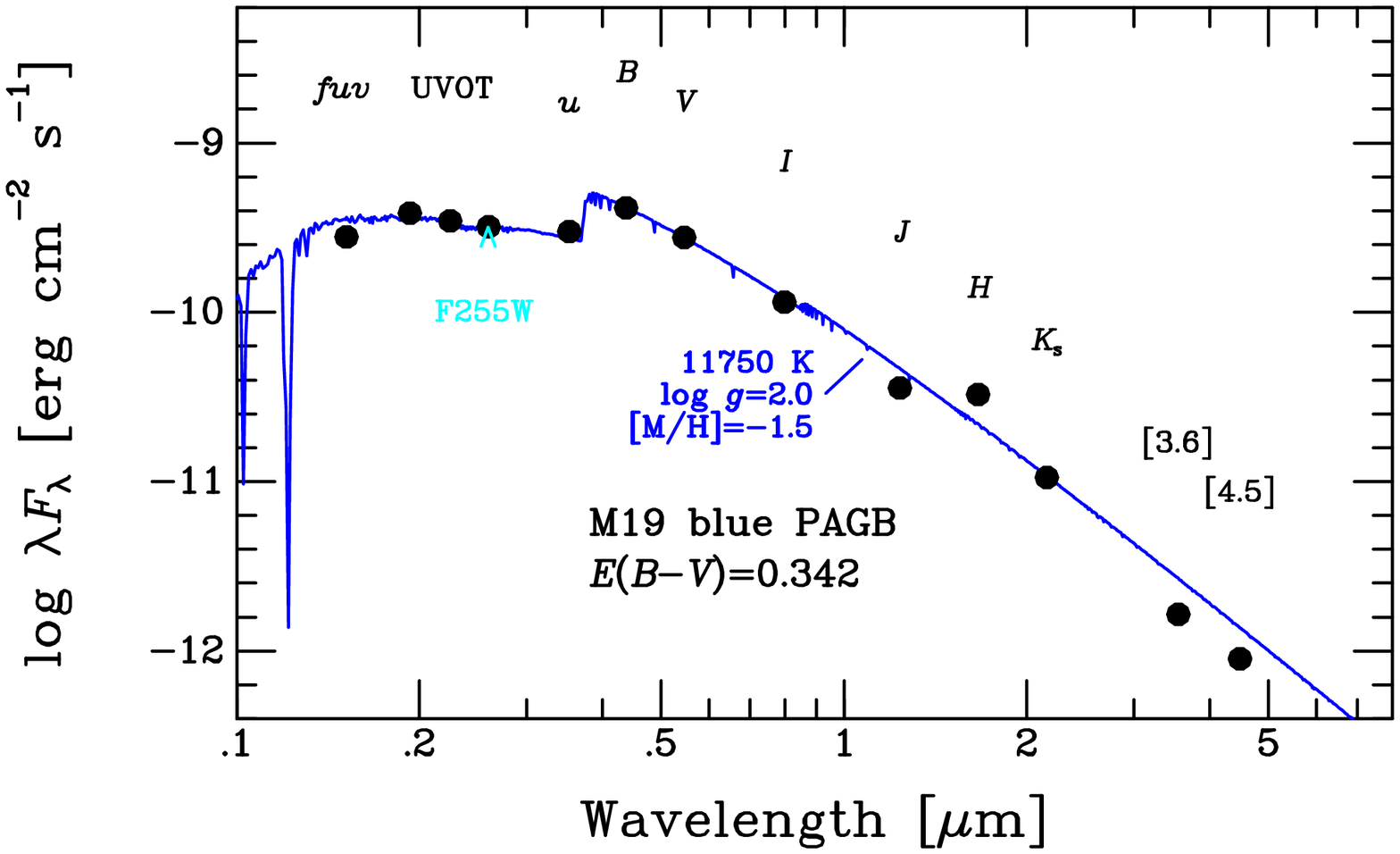}
\figcaption{\footnotesize
Left panel: Spectral-energy distribution for the M19 yellow PAGB star (filled black circles), corrected for interstellar reddening of $E(B-V)=0.344$ as described in the text. The orange curve is a model-atmosphere SED for a star with the parameters indicated in the figure. Note the conspicuously large Balmer discontinuity.
Right panel: SED for the M19 blue PAGB star (filled black circles), corrected for $E(B-V)=0.342$. The \HST\/ observation at F255W (cyan plotting point and label) is a lower limit (see text). The blue curve is a model-atmosphere SED with parameters indicated in the figure. See the text for details of the various public sky surveys used to assemble these SEDs, and caveats about uncertainties due to blending and the extinction corrections.
\label{fig:sed}
}
\end{center}
\end{figure*}

The two panels in Figure~\ref{fig:sed} plot the SEDs for the two PAGB stars. We superpose model-atmosphere SEDs selected from the ATLAS9 grid\footnote{\url{http://wwwuser.oats.inaf.it/castelli/grids.html}} of \citet{Castelli2004}. The metal abundances for these SEDs are $\rm[M/H]=-1.5$, with $\alpha$-element enhancements. We adopted surface gravities of $\log g=0.5$ and 2.0 for the yellow and blue PAGB stars, respectively. The best fits to the SEDs are found for effective temperatures of 6250 and 11750~K, respectively, but we caution that there is some degeneracy between the adopted $\Teff$ values and reddenings.

Using the bolometric corrections for the two Castelli \& Kurucz models, we find the absolute luminosities given in the final row of Table~\ref{table:basicdata}, $\log L/L_\odot=3.24$ and 3.22 for the yellow and blue PAGB stars, respectively. Thus their luminosities are nearly identical, and they are likely to be on very similar post-AGB evolutionary tracks.

A primary aim in investigating the SEDs was to search for evidence of circumstellar dust, ejected during the AGB phase. In general, circumstellar dust has not been found to be prominent in yellow and blue PAGB stars in old populations, such as those in GCs (e.g., B16 and B20), possibly because the PAGB evolutionary timescales are slow enough for any ejecta to have dissipated, combined with the difficulty of forming dust in a low-metallicity environment. As Figure~\ref{fig:sed} shows, we likewise see no evidence for warm circumstellar dust in the two M19 PAGB stars. Unfortunately, the available observations only go out to the \Spitzer\/ [4.5] bandpass; the stellar crowding in the cluster precludes photometry at longer wavelengths with instruments such as \WISE\null. Thus the constraints on dust are not very tight.

\goodbreak

\section{Discussion and Future Studies}

\subsection{Yellow PAGB Stars as Standard Candles}

The main goal of our \uBVI\/ survey was to search the Galactic GC system for yellow PAGB stars and to test them as potential Population~II standard candles. The yellow PAGB star in M19 reported in this paper joins the small number of known objects of this class. We find it to have a visual absolute magnitude of $M_V=-3.39\pm0.09$, based on several recent determinations of the distance to M19. The main contributors to the error are the uncertainties in the cluster's distance, and in the foreground extinction.

B16 listed (their Table~4) the visual absolute magnitudes of the four other known non-variable yellow PAGB stars in Galactic GCs; they range from $M_V=-3.10$ to $-3.46$. Adding our new result for the M19 star, and including also the absolute magnitude for the field yellow PAGB star \bd\ from B20, we find that these six objects have a mean absolute magnitude of $M_V=-3.35\pm0.06$, with a standard deviation of 0.14~mag. This is a very narrow luminosity function, compared for example to that of Population~I Cepheids at a given pulsation period (e.g., \citealt{Clementini2019}). And the discovery and measurement of yellow PAGB stars requires only a single observation epoch. Thus, we believe that these stars continue to be potential extragalactic distance indicators.\footnote{If we had used its yellow PAGB star to estimate the distance to M19, using the zero-point of $M_V=-3.38\pm0.05$ found earlier by B16, we would have obtained $(m-M)_0=14.83\pm0.08$. This agrees extremely well with the mean distance modulus from several independent methods of $(m-M)_0=14.84\pm0.06$, which we discussed in \S\ref{sec:absmag}.} In several forthcoming papers, we will further explore this possibility.

We also note that the five yellow PAGB stars in Galactic GCs all belong to clusters with ``intermediate'' metallicities, of $\rm[Fe/H]=-1.59$ (NGC\,5986), $-1.60$ (M79), $-1.74$ (M19), and $-1.53$ (\oCen), as tabulated by H10. (However, M19 [see \S\ref{sec:observations}] and \oCen\/ have fairly wide ranges of [Fe/H] among their members.) Moreover, all four clusters contain very blue HB stars. We will discuss these and other clues to the evolutionary status of these stars in more detail in papers now in preparation.


\subsection{Future Work}

There are several desirable future investigations of these two new and relatively bright PAGB stars in M19. First, moderate-resolution spectra should be obtained in order to confirm even more definitively that both stars are indeed low-metallicity cluster members, consistent with the interpretations in this paper. High-resolution spectroscopic abundance studies of both of them would be of considerable interest, given the peculiarities seen in other PAGB stars. For example, \citet{Sahin2009} found an anomalously low iron abundance in the yellow PAGB star in M79, and several PAGB stars in younger populations show extreme depletions of refractory chemical elements with high condensation temperatures (see \citealt{Oomen2019} and references therein). Several field PAGB stars appear to be long-period spectroscopic binaries (B20 and references therein), so RV monitoring would be useful to investigate whether binary interactions play a role in the evolution of these objects. Effective temperatures of the stars determined from model-atmosphere fitting, compared to the observed colors, would help refine the estimates of reddening, and thus of their absolute magnitudes. High-precision photometric monitoring would be useful to test whether the yellow PAGB star is a low-amplitude pulsating variable. High-spatial-resolution MIR photometry would provide tighter limits on circumstellar dust than was possible in the study reported here.


\acknowledgments


The observational work in 1998 was partially supported by NASA grant NAG 5-6821 under the ``UV, Visible, and Gravitational Astrophysics Research and Analysis'' program, and by the Director's Discretionary Research Fund at STScI\null. H.E.B. thanks the staff at Cerro Tololo for their support over the years.

This work has made use of data from the European Space Agency (ESA) mission
{\it Gaia\/} (\url{https://www.cosmos.esa.int/gaia}), processed by the {\it Gaia\/}
Data Processing and Analysis Consortium (DPAC,
\url{https://www.cosmos.esa.int/web/gaia/dpac/consortium}). Funding for the DPAC
has been provided by national institutions, in particular the institutions
participating in the {\it Gaia\/} Multilateral Agreement.


Based in part on observations made with the NASA Galaxy Evolution Explorer.
\GALEX\/ was operated for NASA by the California Institute of Technology under NASA
contract NAS 5-98034.  

Based in part on observations made with the NASA\slash ESA {\it Hubble Space Telescope}, obtained from the data archive at the Space Telescope Science Institute. STScI is operated by the Association of Universities for Research in Astronomy, Inc.\ under NASA contract NAS 5-26555.

We acknowledge the use of public data from the Neil Gehrels \Swift\/ Observatory data archive.

This publication makes use of data products from the Two Micron All Sky Survey,
which is a joint project of the University of Massachusetts and the Infrared
Processing and Analysis Center/California Institute of Technology, funded by the
National Aeronautics and Space Administration and the National Science
Foundation.

This work is based in part on observations made with the {\it Spitzer Space Telescope}, operated by the Jet Propulsion Laboratory, California Institute of Technology under a contract with NASA. 

It also makes use of data products from the {\it Wide-field Infrared Survey Explorer}, which is a joint project of the University of California, Los Angeles, and the Jet Propulsion Laboratory/California Institute of Technology, funded by the National Aeronautics and Space Administration.

\facilities{CTIO 0.9m, Gaia, GALEX, 2MASS, Swift, HST, Spitzer, WISE}


\end{document}